\input harvmac


\overfullrule=0pt
\def\Title#1#2{\rightline{#1}\ifx\answ\bigans\nopagenumbers\pageno0\vskip1in
\else\pageno1\vskip.8in\fi \centerline{\titlefont #2}\vskip .5in}

\lref\CDS {A.~Connes, M.~R. Douglas, and A.~Schwarz, {\it
Noncommutative geometry and matrix
theory: Compactification on tori}, { JHEP} {\bf 02} (1998) 003,
{\tt hep-th/9711162}.}

\lref\DH{M.~R. Douglas and C.~Hull, {\it
D-branes and the noncommutative torus},
{ JHEP} {\bf 02} (1998) 008,
{\tt hep-th/9711165}. }

\lref\malda{J.~Maldacena, {\it
The Large N limit of superconformal field theories and
supergravity}, {Adv. Theor. Math. Phys.} {\bf 2} (1998) 231,
{\tt hep-th/9711200}.}

\lref\RT{J.G. Russo and A.A. Tseytlin, {\it Waves, boosted
branes and BPS states in M-theory}, Nucl.Phys.B490 (1997) 121,
{\tt hep-th/9611047}.}

\lref\GH{S. Gubser and A. Hashimoto,
{\it Exact absorption probabilities for the D3 brane},
Commun.Math.Phys. 203 (1999) 325 {\tt
hep-th/9805140}.}

\lref\bergshoeff{ E. Bershoeff, C. Hull and T. Ort\'\i n,
Nucl. Phys. B451 (1995) 547,
{\tt hep-th/9504081}.}

\lref\GKP{S. Gubser, I. Klebanov and A. Polyakov, {\it Gauge theory
correlators from non-critical string theory}, Phys.Lett. B428 (1998) 105,
{\tt hep-th/9802109}.}

\lref\filk{T. Filk, {\it Divergences in a field theory on quantum space},
Phys.Lett. B376 (1996) 53;
T. Krajewski and R. Wulkenhaar, {\it 
Perturbative quantum gauge fields on the non-commutative torus},
{\tt hep-th/9903187}; D. Bigatti and L. Susskind, 
{\it Magnetic fields, branes and noncommutative geometry},
{\tt hep-th/9908056}.}

\lref\witten{E. Witten, {\it Anti-de Sitter space and holography},
{Adv. Theor. Math. Phys.} {\bf 2} (1998) 253, {\tt hep-th/9802150 }.}

\lref\sw{ N. Seiberg and E. Witten, {\it String Theory and Noncommutative
Geometry}, 
hep-th/9908142. }

\lref\myers{J. Breckenridge, G. Michaud and R.C. Myers, {\it
More D-brane bound states}, Phys.Rev. D55 (1997) 6438,
{\tt hep-th/9611174}.}

\lref\luroy{J.X. Lu and S. Roy, {\it ((F,D1),D3) bound state and its
T-dual daughters}, {\tt hep-th/9905014}.}

\lref\maldatwo{N. Itzhaki, J. Maldacena, J. Sonnenschein and
S. Yankielowicz,
{\it Supergravity and the large N limit of theories with sixteen
supercharges},
Phys. Rev. {\bf D58} (1998) 046004, {\tt hep-th/9802042}.}

\lref\genref{
F. Ardalan, H. Arfaei and M.M. Sheikh-Jabbari, 
{\it noncommutative geometry from strings and branes}, JHEP 9902 (1999) 016,
{\tt hep-th/9810072}; 
M. Sheikh-Jabbari, {\it 
Super Yang-Mills Theory on Noncommutative Torus from Open Strings Interactions
}, Phys.Lett. B450 (1999) 119, {\tt hep-th/9810179}.
} 

\lref\ABS{
O. Aharony, M. Berkooz and Nathan Seiberg,
{\it Light cone description of (2,0) superconformal theories in 
six-dimensions}, 
Adv.Theor.Math.Phys.2 (1998) 119, 
{\tt hep-th/9712117}.}

\lref\IH{
A. Hashimoto and N. Itzhaki, 
{\it Noncommutative Yang-Mills and the AdS/CFT correspondence},
{\tt hep-th/9907166}.} 

\lref\witt{E.~Witten, {\it Anti-de Sitter space,
thermal phase transition and confinement in gauge theories},
{Adv. Theor. Math. Phys.} {\bf 2} (1998) 284, {\tt hep-th/9803131 }.}

\lref\landau{L.D.~Landau and E.M.~Lifshitz, {\it Quantum Mechanics}.}

\lref\shol{O. Aharony, M. Berkooz, D. Kutasov and 
N. Seiberg, {\it Linear Dilatons, NS fivebranes and Holography},
hep-th/9808149; S. Minwalla and N. Seiberg, {\it Comments on the IIA 
(NS) fivebrane}, JHEP 9906:007, 1999, hep-th/9903142.}

\lref\HS{ G. Horowitz and A. Strominger, {\it
Black Strings and $p$-branes}, Nucl. Phys. {\bf B360} (1991) 197.}

\lref\BFSS{ T. Banks, W. Fischler, S. Shenker and L. Susskind,
{\it M Theory as a Matrix Model: A Conjecture}, 
Phys. Rev. {\bf D55}(1997) 5112, hep-th/9610043.}


\Title{\vbox{\baselineskip12pt \hbox{hep-th/9908134}
\hbox{HUTP-99/A046 }}}
{\vbox{\centerline {Large $N$ Limit
of Non-Commutative Gauge Theories }
}}
%
%
\centerline{Juan M. Maldacena~$^{a}$\ and \ Jorge G. Russo~$^{b}$ }

\bigskip
\centerline{$^{a}$ {\it Department of Physics,
Harvard University} }
\centerline{\it Cambridge, MA 02138}

\medskip
\centerline{$^{b}$ {\it Departamento de F\'\i sica,
Universidad de Buenos Aires} }
\centerline{\it Ciudad Universitaria, 1428 Buenos Aires}

%
%

\def\[{\left [}
\def\]{\right ]}
\def\({\left (}
\def\){\right )}

\def\p{\partial}

\vskip .3in

\centerline{\bf Abstract}

Using the correspondence between gauge theories and
string theory in curved backgrounds, we investigate
aspects of the large $N$ limit of non-commutative gauge theories
by considering gravity solutions with $B$ fields.
We argue that the total number of physical 
degrees of freedom at any given scale coincides with the commutative case.
We then compute a two-point correlation function 
involving momentum components in the directions of the $B$-field.
In the UV regime, we find that the two-point function decays exponentially 
with the momentum. 
A calculation of Wilson lines
suggests that strings cannot be localized near the boundary. We also
find string configurations that are localized in a finite region 
of the radial direction. These are worldsheet instantons.


\smallskip
\noindent
\Date{}


\newsec{Introduction}

Gauge theories on non-commutative spaces can arise in certain limits of 
string theory \refs{\CDS, \DH ,\genref , \sw }.
Specifically, one considers a system of D$p$-branes with 
a constant NS $B$ along their worldvolume directions. In general,
there are
open strings and closed strings coupled together, but it is 
possible to take a low energy limit --rescaling some parameters of the 
theory, like the metric along the worldvolume directions and the string
coupling constant-- in such a way that the closed strings decouple
and the resulting action for the open string modes is 
the one corresponding to a 
non-commutative gauge theory \refs{\CDS,\sw}.

In this paper we consider gravity solutions that are related to D$p$ branes
with $B$ fields. In the spirit of \refs{\malda,\maldatwo},
the near horizon region of these
gravity solutions should describe the large $N$
limit of non-commutative gauge theories. We will 
see that the limit 
we have to take in order to isolate the near horizon region is the same
as the limit that is taken from the field theory point of view.
These gravity solutions reduce to the usual D$p$ brane solutions 
very close to the horizon; this corresponds to the fact that the 
field theory reduces to the commutative field theory at long distances. 
In all cases the dilaton 
goes to zero at the boundary faster than in the commutative counterparts.
Close to the boundary the solution is quite different from the usual
D$p$ brane solution and it typically contain a varying $H=dB$ NS field
as well as some RR field strengths.
The presence of a $B$ field induces a D$(p-2)$ charge density. 
In fact, we can gauge away $B$ in the bulk at the expense of introducing 
a $U(1)$ field strength on the brane theory. So some of 
the gravity solutions 
are essentially the ones found in \refs{\RT,\myers }.

Our discussion will mostly concentrate 
on solutions describing $D3$ branes with 
$B$ fields. 
We will see that correlation functions between operators
with vanishing momentum along the directions 
with non-zero $B$ field give the same 
correlation functions as in the case with zero $B$-field. This is in 
agreement with the arguments in \filk , which indicate
that planar diagrams depend on the non-commutativity
parameter only through the 
external momenta. Correlators with dependence on the momenta 
along the $B$ directions are
quite different. In particular, we find that
the equation for small fluctuations of a 
certain graviton mode is formally
the same as the wave equation in the full
D3-brane geometry. 
The renormalization factors necessary to define the correlators depend on
the momenta. This renders two-point functions rather 
ambiguous, since one can substract any function of the external momenta.
(The ratio of three point functions to two point functions should be 
unambiguous). 
We discuss a prescription to define correlation functions, 
and calculate the two-point function corresponding to
the components of energy-momentum tensor of the gauge theory.
We choose our prescription so that  at
 low energies the correlator reproduces the usual
 $k^4\log k^2 $ behavior of the standard commutative gauge theory.
At high energies, we find an exponential fall off with the
momentum.

By the holographic 
principle,
one can determine 
the number of degrees of freedom 
by computing the area of surfaces at fixed $r$, and relating them
to the number of degrees of freedom of the field theory with an energy 
cutoff of order $r$. The number of degrees of freedom, computed in this
fashion, is the same as in the commutative case,
though there seems to be redistribution of them, in a sense that
will be explained.


The supergravity solution in Minkowski space with 
a $B$ field in the time direction is also given. 
This corresponds to a D3 brane with 
electric and magnetic U(1) worldvolume fields. 
Other generalizations are discussed in section 6.
We give gravity solutions corresponding to M5 branes in the
presence of 
a constant 
$C_{\mu\nu\rho}$ field --~a 
theory whose DLCQ version was studied in 
\ABS ~-- and we construct solutions for D1-D5 systems with $B$ fields.
These include a solution where
the $B$ field is along the worldvolume of the D1 brane, and
the case where the $B$ field is on a torus inside the 5-brane
world volume.

While this paper was being written we received the paper \IH\ which
has some overlap with section 2.

\newsec{Construction of the solutions}

\subsec{D3 brane in constant $B$ field}

\def\ap{{\alpha'} }

In this section we obtain the solution of a D3 brane in a constant
NS $B$ field background.
All solutions with non vanishing $B$ fields in this section preserve
16 supersymmetries. First we consider the case where there is only one
component $B_{23}$
different from zero.
A simple way to obtain the solution is
to perform a T-duality along $x_3$. This gives a smeared D2 brane on
a tilted torus. It is easy to write down this solution. Then we T-dualize
back on $x_3$, using the T-duality rules for RR backgrounds derived in
\bergshoeff . The solution in string metric is
\eqn\solmet{\eqalign{
ds^2_{str} &= f^{-1/2} [ - dx_0^2 + dx_1^2 +{h}(dx_2^2 + dx_3^2 )]
+ f^{1/2} ( dr^2 + r^2 d \Omega^2_5 )\ ,
\cr
f = & 1 + {\ap ^2 R^4\over r^4 } ~, ~~~~~~~~~
h^{-1} =\sin^2\theta f^{-1}+\cos^2\theta \ ,
\cr
B_{23} =& {\sin\theta\over \cos\theta }\ \ f^{-1} h \ ,
\cr
e^{2\phi} &= g^2 h \ ,
\cr
F_{01r} &= {1 \over g} \sin\theta\ \partial_r f^{-1} \ ,
\ \ \ \ \
F_{0123r} = {1 \over g} \cos\theta \ h \ \partial_r f^{-1} \ .
}}
The asymptotic value of the $B$ field is $B^{\infty}_{23} = \tan \theta $.
The parameter $R$ is defined by
\eqn\defr{
\cos\theta \ R^4 = 4 \pi g N \ ,
}
where $N$ is the number of $D3$ branes, and $g \equiv g_{\infty}$ is the
asymptotic value of the coupling constant.
It is possible to gauge away a constant
$B$-field. However, this introduces a constant flux for the worldvolume
gauge
field, since under $B \to B + d \Lambda$, $A \to A - \Lambda$.
In fact, performing this gauge transformation we obtain a solution which
is the same as \solmet\ except that the value of the $B$ field is shifted
by $B \to B - B^{\infty} $. 
In that form, it is easy to see that the solution
has D3 and D1 brane charges. This solution was found in 
\refs{\RT,\myers },
and it represents D1 branes dissolved
in D3 branes. 

%

In Euclidean space, one can also construct a solution with
$B_{01} = \tan \theta'$, $ B_{23} =\tan \theta $
in a similar way.
The solution in string metric is given by
\eqn\smb{\eqalign{
ds^2_{str} &= f^{-{1/2}} [ h'
(dx_0^2 + dx_1^2 ) + h (dx_2^2 + dx_3^2) ]
+ f^{1/2}[ dr^2 + r^2 d \Omega^2_5 ]\ ,
\cr
f & = 1 + {\ap ^2R^4 \over r^4 } ~, ~~~~~~~~~
h^{-1} =\sin^2\theta f^{-1}+\cos^2\theta \
~,~~~~~~ {h'}^{-1}=\sin^2\theta' f^{-1}+\cos^2\theta' \ ,
\cr
B_{01} &= {\sin\theta'\over \cos\theta' }\ f^{-1} h' \ ,\ \ \ \ \ \
B_{23} = {\sin\theta\over \cos\theta }\ \ f^{-1} h \ ,
\cr
e^{2\phi} &= g^2 h h'\ ,\ \ \ \ \
\chi = i {1 \over g}\sin\theta\sin\theta ' f^{-1}\ ,
\cr
A_{01} &= i{1 \over g} \sin\theta \cos\theta' \ h' f^{-1} \ ,\ \ \ \ \
A_{23} = i{1 \over g} \sin\theta '\cos\theta \ h f^{-1}\ ,
\cr
F_{0123r} & = i{1 \over g} \cos\theta\cos\theta ' h h' \partial_r f^{-1}\ .
}}
The parameter $R$ is given by
$ R^4 = 4 \pi g N (\cos\theta \cos\theta')^{-1} $.

There is an analogous solution with
Lorentzian signature which is obtained
by a Wick rotation $x_0\to ix_0$ and $\theta'\to i\theta '$
(so that $\cos\theta '\to \cosh\theta '$, $\sin\theta '\to i\sinh\theta' $).
As a result, the imaginary factors in the gauge fields
disappear.
The Lorentzian solution can also be obtained from eleven dimensions
by starting with a stack of
M2-branes (with world-volume coordinates $x_0,x_1,x_2$~)
smeared in two direcions $x_3,x_4$.
One redefines coordinates as follows:
\eqn\aaa{ \eqalign{
x_4&= \tilde x_4 \cos\alpha+ (\tilde x_2 \cos\theta+ \tilde
x_3\sin\theta )
\sin\alpha\ ,
\ \ \ \
\cr
x_2&=- \tilde x_4 \sin\alpha+ (\tilde x_2 \cos\theta+ \tilde
x_3\sin\theta )\cos\alpha\ ,
\cr
x_3&=-\tilde x_2 \sin\theta+ \tilde x_3\cos\theta\ .
}}
Next, we make dimensional reduction in $ \tilde x_4$
and a T-duality transformation in the $\tilde x_3$-direction.
We find
\eqn\fff{
\eqalign{
ds^2_{str} &=f^{-1/2}\big[ h'(-dx_0^2+dx_1^2)+
h(dx_2^2+dx_3^2)\big]+f^{1/2}
\big[dr^2+r^2d\Omega_5^2\big]\ ,\
\cr
f &=1+\cos^2\alpha {\ap ^2R^4_0\over r^4}\ ,\ \ \ \ \
h={f\over G}\ ,\ \ \ \ \ h'={f\over H}\ ,
\cr
H &=1+{\ap ^2R_0^4\over r^4}\ ,\ \ \ \ \ \ G =1+\cos^2\alpha\cos^2\theta \
{\ap ^2 R^4_0\over r^4}\ ,\ \
\cr
e^{2\phi } & = g^2 hh'\ .
}}
The function $H$ is the harmonic function appearing in the original M2 brane
metric. The solution is indeed the Wick rotated version
of \smb , with the identification of parameters:
$ R^4_0 =R^4 \cosh^2\theta ' $, $\cos^2\alpha =1/\cosh^2\theta ' $.
The particular case $\theta =0 $ appeared 
in \luroy . In this case the background \fff\
reduces to the solution which is obtained by
applying $S$-duality transformation to the solution \solmet .

\subsec{Decoupling limit}

\def\ap{ {\alpha '} }

The above solutions are asymptotic to flat space for $r \to \infty$ and
they have a horizon at $r =0$. Very near $r =0$ the solutions look like
$AdS_5 \times S^5$. The throat region connecting these two contains
non-zero NS and RR $B$ fields.
If we take the standard low energy limit, keeping all other parameters
constant we just get the usual $AdS$ solution. On the boundary we have
the usual ${\cal N}=4$ Yang-Mills theory. In order to obtain non-commutative
Yang-Mills we should also take the $B$ field to infinity.
The supergravity solution itself tells us what limit we should take
to decouple the asymptotic region while still keeping the region of
the spacetime where $B$ fields vary.

For the solution \solmet\ the rescaling of the parameters should be the
following
\eqn\limits{\eqalign{
& \alpha' \to 0 ~,~~~
\tan \theta = {\tilde b \over \alpha'}~,~~~\cr
& x_{0, 1} = \tilde x_{0,1} ~,~~~~
x_{2,3} = {\alpha' \over \tilde b} \tilde x_{2,3} ~, \cr
& r = \alpha' R^2 u ~,~~~~
g = {\alpha'\over \tilde b} \hat g\ ,
}}
where $\tilde b, ~ u, ~ \hat g, ~ \tilde x_\mu$ stay fixed.
This scaling of parameters is precisely the same as
the scaling that
was found in \sw . The factor of $\tilde b$ in the second line of \limits\
is introduced just for later convenience.

\eqn\solnear{\eqalign{
ds^2_{str} &= \alpha' R^2 \bigg[ u^2 ( - d {\tilde x}_0^2 +
d{\tilde x}_1^2)+
u^2 \hat h (d{\tilde x}_2^2 + d {\tilde x}_3^2 )
+ {du^2\over u^2} + d \Omega^2_5 \bigg]\ ,
\cr
\hat h &= { 1 \over 1 + a^4 u^4} ~,~~~~~~~~~ a^2 =
{\tilde b R^2} \ ,
\cr
\tilde B_{23} &=
B_\infty {a^4 u^4 \over 1+a^4u^4}\ ,\ \ \ B_\infty=\alpha '{ 1\over \tilde
b} = \alpha' {R^2 \over a^2}
\ ,
\cr
e^{2\phi} &= {\hat g}^2 \hat h \ ,
\cr
A_{01} &= {\ap } {\tilde b\over \hat g} {u^4}R^4 ,
\cr
\tilde F_{0123u} & ={\ap }^2{\hat h \over \hat g}
\partial_u ( u^4 R^4)\ ,
}}
where $\hat g $ is the value of the string coupling
in the IR. Now we have $R^4 = 4 \pi \hat g N$. Notice
the $B$ and $A$ fields are defined with respect to
the new coordinates and they include some rescalings implied by \limits .
We propose that \solnear\ is the dual gravity description of
non-commutative
Yang-Mills.
This solution reduces to the $AdS_5 \times S^5$
solution for small $u$, which corresponds to the IR regime of the gauge
theory.
This is consistent with the expectation that non-commutative Yang-Mills
reduces
to ordinary Yang-Mills theory at long distances.
The parameter $a$ has the dimension of length, and
the solution starts deviating from the $AdS_5 \times S^5$ solution
at $u \sim 1/a$, i.e. at a distance scale
of the order $a = R \sqrt{\tilde b} $.
For large $ \hat g N \sim R^4$, this is greater
than
the naively expected\foot{
{}From the limiting form of the two point function of the coordinate
of the boundary of an open string we find that $[\tilde x^2 , \tilde x^3]
\sim \tilde b $ \refs{\CDS ,\sw } .} distance scale of 
$L \sim \sqrt{\tilde b}$.
This can be interpreted as due to strong interactions, which seem
to render visible
the effects of non-commutativity at longer distances than
naively expected.
The solution has a boundary at $u = \infty$.
As we approach the boundary, the physical size
of the $2,3$ direction shrinks (in string units), since
$\hat h \sim 1/u^4 $ for large $u$.
All curvature invariants in string metric remain bounded as we approach the
boundary. The reason for
this
is that the string metric near the boundary has a scaling isometry:
$ x_{0,1} \to \lambda^{-1} x_{0,1}, ~ u \to \lambda u,~
x_{2,3} \to \lambda x_{2,3}$. Consequently, curvature invariants
of the metric near $u=\infty $ (obtained by $\hat h\to 1/(au)^4$~) cannot
depend
on $u$, for $u \to \infty$.
However, if we were to compactify the $2,3$ directions we would get a
singularity since some winding modes would become light.
This singularity cannot be avoided by going to the T-dual description. In
fact,
T-duality transformations along
$x_2,x_3$ lead to the same background \solmet\ with the exchange
$\cos\theta \leftrightarrow \sin\theta $.
In the case that the $B$ field is rational the singularity can be
removed by performing an appropriate $SL(2,Z)$ T-duality transformation, 
as explained in \CDS . These decoupled solutions still have 16
supersymmetries,
there is no supersymmetry enhancement since the theory is not conformal.

The near-horizon background \solnear\ can also be obtained from eleven
dimensions
by starting from the standard M2 brane solution smeared in two directions $x_3,x_4$,
making dimensional reduction in a null circle $x^+= x_4+x_0$,
and T-duality in $ x_3$. 
This is related to the observation of \CDS\ that M theory compactified on a null circle with non-zero $C_{+ij}$ leads to non-commutative geometry.
Thus the solution \solnear\ has also the eleven-dimensional interpretation
of an M2 brane with an infinite boost in a transverse direction.

Let us now consider the decoupling limit of the geometry \smb . The
rescaling of the parameters is basically the same as \limits\ except that
now one should also rescale the $0,1$ coordinates
and the string coupling
in a slightly different way,
\eqn\limitstwo{
g = {\alpha'^2\over \tilde b\tilde b'} \hat g ~,~~~ x_{0,1} = 
{\alpha' \over \tilde b'} \tilde x_{0,1}\ \ .
}
We introduce parameters $a,a'$
with dimension of length by $a^2 = \tilde b R^2,\ {a'}^2= \tilde b'R^2 $.
We get (after relabelling $\tilde x_i\to x_i$ of the world-volume coordinates)
\eqn\snn{\eqalign{
ds^2 = & \alpha 'R^2 \bigg[ u^2\big[ \hat h' ( dx_0^2 + dx_1^2)+
\hat h (d{ x}_2^2 +
d { x}_3^2 ) \big]
+ {du^2\over u^2} + d \Omega^2_5 \bigg]\ ,
\cr
e^{2\phi}= & {\hat g}^2 \hat h \hat h'\ , \ \ \ \
\hat h = { 1 \over 1 + a^4 u^4} ~,~~~~\hat h ' = { 1 \over 1 + {a'}^4 u^4} \
,
\cr
B_{01}= & \alpha ' R^2 {{a'}^2u^4 \over 1+ {a'}^4 u^4}\ ,\ \ \ \ \
B_{23}=\alpha ' R^2 {{a}^2u^4 \over 1+ {a}^4 u^4}\ ,
\cr
A_{01} = & i \alpha' { \tilde b \over \hat g} \hat {h'} R^4 u^4\ , \ \ \ \ \
A_{23} = i \alpha' { \tilde {b'} \over \hat g} \hat {h} R^4 u^4 \ ,
\cr
F_{0123u} = & i \ap ^2 { \hat h \hat h'\over \hat g} \partial_u ( R^4
u^4) \ , \
\ \ \ \ \
\chi = i { \tilde b \tilde {b'} \over \hat g} { R^4 u^4 }\ .
}}
where again $\hat g $ is  the value of
the
string coupling in the IR.
This solution reduces to the standard $AdS_5 \times S^5 $ solution for
small $u$. The effects of non-commutativity appear at the largest of the
two scales $a=R\sqrt{\tilde b}$ or $a'=R\sqrt{\tilde{b'} }$.
For large $u$ the physical size of the spatial directions decreases, but
all curvature invariants in string metric are bounded. In fact, the string
metric for large $u$ is again $AdS_5 \times S^5$.
Although the point $u = \infty $ may look like the ``horizon'' of $AdS$,
the other fields are different from the fields of the usual $AdS_5 \times
S^5 $ solution.
In particular,
the dilaton $e^{\phi}$ goes to zero at large $u$.
This implies that fluctuations
close to the boundary are suppressed (that is why we call it
``boundary''). This boundary is similar in spirit to the
case of the NS fivebrane. There also we find that the metric in string
units
remains constant as we move away from the 
horizon. For the NS fivebrane the
dilaton goes to zero and this suppresses interactions as we approach the
boundary.
The fact that the dilaton goes to zero also freezes the asymptotic
geometry,
which is the physical property that we expect from a boundary.
A completely analogous situation arises in the near horizon geometry of
D-instantons. There again the dilaton goes to zero and the physical
size in string metric goes to zero. In fact, the D-instanton boundary
looks like the origin of $R^{10}$ except that the string coupling
is going to zero. This similarity is not a coincidence, in fact the
behavior of the solution \snn\ close to $u \sim \infty$ is
the same as the behavior of a D-instanton smeared in the 0123 directions.
If we compute the area in Planck units of a surface of constant $u$
we see that it increases as $u^4$. This is the same behavior that
one finds in the pure commutative Yang-Mills.

The decoupling limit for the metric \smb\ with Lorentzian signature
has a different form, since the limit
must be taken in a different way. As it is clear from the
eleven-dimensional
origin as a stack of rotated M2 branes \fff ,
the appropriate limit now amounts to dropping the
``1" in the harmonic function $H$.
The string coupling $e^\phi $ becomes strong at large $u$.
Remember that we find the Lorentzian signature version of \smb\ by
replacing $\theta' \to i \theta'$. 
The scaling of parameters is as follows:
\eqn\limm{\eqalign{
&\cosh \theta'  = {  \tilde b' \over \alpha'}\ ,\ \ \ \ \ \
\cos\theta = fixed,
\cr
& x_{0, 1} =  {\tilde b' \over \sqrt{ \alpha'}} \tilde x_{0,1} ~,~~~~
x_{2,3} = \sqrt{\alpha' } \cos\theta \tilde x_{2,3} ~, \cr
& r = \sqrt{\alpha'}  u ~,~~~~~
g = {\tilde b' \cos\theta \over \alpha' } \hat g\ 
\cr
 & R^4  = fixed = 4 \pi N \hat g.
}}
We get
\eqn\slor{\eqalign{
ds^2 = & \alpha ' f^{1/2} \bigg[ { u^4 \over R^4}(- d{\tilde x_0}^2 +
 d{\tilde x_1}^2) + { u^4 \over R^4}\hat h (
d{\tilde x_2}^2 +
 d{\tilde x_3}^2) + du^2 + u^2 d\Omega_5^2 \bigg]
 \cr
f & = 1 + {R^4 \over u^4 } ~, ~~~~~~~~~
{\hat h}^{-1} = 1 + { u^4 \over R^4 \cos^2\theta } \ ,
\cr
B_{01} &= \alpha' { u^4 \over R^4}  \ ,\ \ \ \ \ \
B_{23} =   \alpha' \tan\theta { u^4 \over R^4} \hat h \ ,
\cr
e^{2\phi} &= {\hat g}^2 f^2 {u^8 \over R^8} \hat h\ ,\ \ \ \ \
\chi = { 1 \over \hat g } \tan \theta f^{-1}\ ,
\cr
A_{01} &= \alpha' {1 \over \hat g} \tan\theta {u^4 \over R^4} \ , \ \ \ \ \
A_{23} = \alpha' {1 \over \hat g}{u^4 \over R^4}{\hat h} \ ,
\cr
F_{0123u} & = {\alpha'}^2 {1 \over \hat g} \hat h { 4 u^3 \over R^4 } \ .
}}

\newsec{Generalization to non-extremal metrics}

\subsec{Non-commutative gauge theory at finite temperature}

A B-field can be similarly introduced in the
non-extremal D3 brane background
by the same U-duality transformations that
were made above for the extremal case.
The resulting metric is simply obtained by
multiplying the $g_{00}$ component by
$(1-{u_h^4\over u^4})$, and the $g_{uu}$ component by
$(1-{u_h^4\over u^4})^{-1}$.
In particular, the non-extremal version of the solution \solnear\
in the Einstein frame $ds^2_{E}=e^{-\phi/2} ds^2_{str}$ is given by
\eqn\eei{ \eqalign{
ds^2_E &= R^2 (1+a^4u^4)^{1/4} \bigg( u^2 \big[
-(1-{ u_h^4\over u^4}) dx_0^2+dx_1^2+{1\over (1+a^4u^4)}(dx_2^2+dx_3^2)\big]
\cr
&+ \ {du^2\over u^2(1-{u_h^4\over u^4}) }+ d\Omega_5^2\bigg)\ .
}}
String theory on this non-extremal background should provide a dual
description of
non-commutative Yang-Mills theory at finite temperature.

An interesting question concerns
the number of degrees of freedom of non-commutative gauge theories.
Because of the non-commutative nature of the space-time coordinates,
one may expect
that in the ultraviolet regime there could be
a reduction of the degrees of freedom relative to the usual gauge theories.
The number of gauge-invariant degrees of freedom is reflected on
the dependence of the free energy with the temperature.
Let us denote by $S_0, E_0$ and $F_0$ the entropy, energy and free energy
of the black D3-brane with $B_{23}=0$ ($a=0$). The Hawking temperature of the
metric \eei\ is clearly the same as the temperature of the $a=0$ metric,
since the $u,x_0$ part of the metric is only modified by a conformal factor.
The entropy $S$ for the metric \eei\ is
determined by the area of the event horizon times the volume of the
D3-brane, that is, $S={V_8\over V_8^0}S_0$, where $V_8$ is the volume
of the $(x_1,x_2,x_3,\Omega_5)$ space, and $V_8^0$ the
corresponding volume for the metric with $a=0$.
{}From eq.~\eei , it follows that $V_8=V_8^0$, and therefore
$S=S_0$. 
By the first law of thermodynamics, $dE=T_HdS$, one can anticipate that 
the energy is also unchanged. In fact, it is easy to see that this is the
case by calculating the mass directly. Thus $F=E-T_H S=F_0$.
The fact that all thermodynamic quantities are the same as in 
ordinary gauge theories is indicating that the number of degrees of 
freedom is also the same.
The same conclusion applies for the other solutions discussed in this paper.
This is consistent with the arguments of \filk , 
saying that all large $N$ computations
with no external lines should give the same result, 
which is what we find.

Although the total number of degrees of freedom remains
unchanged relative to the usual case as the energy scale is varied,
there seems to be a redistribution of them.
As $u\to\infty $, a contraction of the volume of the torus $x_2,x_3$ is
compensated by an expansion of the volume of the sphere.
This means that momentum modes become heavier while angular modes become
lighter.

\subsec{Non-commutative Models with ${\cal N}=0$ supersymmetries}

Let us now consider the system \snn\ with $B_{01}$ and $B_{23}$
in the particular case $a=a'$.
The generalization to the non-extremal case is easy if one
starts with the usual
non-extremal M2 brane metric, and make dimensional
reduction and T-duality as described in \aaa , and finally
the Wick rotation to Euclidean signature. We then
take the decoupling limit as before.\foot{As explained above,
the decoupling limit on the Euclidean solution is not equivalent
to the decoupling limit of the Lorentzian solution \slor .}
The near-horizon, non-extremal metric 
in the Einstein frame is given by
\eqn\mko{ \eqalign{
ds^2_E &=\alpha ' {R^2a^2\over \sqrt{\hat g}} 
\bigg( \tilde f^{-1/2}\big[\big(1-
{u_h^4\over u^4}) dx_0^2+dx_1^2+dx_2^2+dx_3^2\big]
\cr
&+\ \tilde f^{1/2}\big[ {du^2\over 1-{u_h^4\over u^4}}
+u^2d\Omega_5^2\big]\bigg)\ ,
}}
$$
e^{2 \phi} = {\hat g}^2 (1 + a^4 u^4)^{-2}~,~~~~~~~~
\tilde f=1+ {1\over a^4u^4}\ .\ 
$$
(We have rescaled $x_i \to a^2 x_i $, $i =0,...,3$).
Remarkably, the metric \mko\ coincides with the euclidean metric
of the black D3 brane. The full background is of course not equivalent,
since other fields are also varying and we have written the dilaton
as an example.
The identification of parameters with the standard non-extremal solution of \HS\ is as follows: $a=1/r_-$,
$u_h^4=r_+^4-r_-^4$, where $r_-,\ r_+$ are the inner and outer
horizons in coordinates $r^4=u^4 +r^4_-$.
The coordinate $x_0$ is periodic and describes a circle of 
radius $\rho_0$, given by
\eqn\hhgg{
\rho_0={1\over 2\pi T_H}={ r_+^2\over 2} (r_+^4-r_-^4)^{-1/4}
={1\over 2u_h} \sqrt{u_h^4+{1\over a^4}}\ .
}
The D3 brane charge of the usual \HS\ black D3 brane solution
is related to $r_\pm $ by
$r_+^2r_-^2=4\pi g N\ap ^2 $. 
In the present case, the charge $N$ (which is in $R^4=4\pi\hat g N$)
is not related to $r_+, r_-$;
now $r_-=1/a$ and $r_+$ is a function of $a$ and the 
radius $\rho_0$. 
String theory on the background \mko\
should be related to the euclidean non-commutative $SU(N)$ gauge theory 
with $B_{01}=B_{23}$ compactified, with antiperiodic boundary 
conditions, on the circle
parametrized by $x_0$.

The usual black D3 brane metric \HS\ has rather non-trivial 
thermodynamic properties,\foot{
At a certain temperature $T_{\rm cr}$ (corresponding to
$r_+^4/r_-^4=\sqrt{3}$ or 
$M/M_{\rm ext}=7/2\sqrt{3}$) the specific heat becomes infinity, 
and above that temperature it becomes negative, 
whereas,
for the solution with $\tilde f\to {r_-^4 \over u^4}$,
the specific heat is constant and positive.}
and one can expect that this will translate into some 
interesting effect in the gauge theory dual to \mko .
%
%
%
The parameter $u_h$ is given in terms of $\rho_0$ and $a$ by \hhgg , i.e.
$u_h^2=2\rho_0^2\pm\sqrt{4\rho_0^4-1/a^4}$.
There is a minimum radius $\rho_0^2=\rho_{\rm min}^2=1/(2a^2) $,
below which we cannot find smooth solutions of this kind
(in terms of the original coordinates $x_i$ with dimension of length,
obtained by $x_i\to x_i/a^2$, one has
$\rho_{\rm min}= a/\sqrt{2}$).
For every $\rho_0$ such that $\rho_0\geq \rho_{\rm min}$ ,
there are two possible values of $u_h$.
The one which is connected to the $AdS_5$ solution is that
of smaller $u_h$.
The branch with greater $u_h$ is unstable.
The reason is that it is essentially the same as a non-extremal
Schwarzschild black hole in 7 dimensions.

The antiperiodic boundary condition
gives rise to tree-level masses of $O(1/\rho_0)$
for all fermions and supersymmetry
is completely broken to ${\cal N}=0$.
In the present case, the radius cannot be taken to zero,
since as explained above
the metric cannot be smooth at $u=u_h$ for any
$\rho_0< \rho_{\rm min }$.
It is worth noting that the spectrum of the Laplace operator
corresponding to the Einstein-frame metric \mko\ is continuous.
The reason is that there are plane waves at infinity, where the
metric is flat.
This suggests that the mass spectrum of physical fluctuations is continuous
(modulo possible subtleties about boundary conditions due to the varying
dilaton and other fields).
Clearly, more investigation on this model
is needed.

\newsec{Correlation functions}

Let us first consider correlation functions in the case that only
$B_{23} \not =0$. If we consider fields that do not depend on $x_{2,3}$
then we can easily see that their correlators give the same result
as 
in $AdS_5 \times S^5$ in the same
situation,
i.e. zero momentum in the 2,3 directions. The reason is that we obtained
the solution by a U-duality transformation that involved the $2,3$
directions.
Since this is a continous symmetry of the gravity solution --regardless of
whether the coordinates are compact or not-- any given boundary
conditions for the fields can be translated into boundary conditions
in $AdS_5 \times S^5$; one can then calculate the perturbed
supergravity solution, which
by asumption will not depend on $x_2,x_3$, and T-dualize it
back to obtain the correlation functions.

Non-trivial correlations will involve situations where the fields
depend on $x_2, x_3$. Now both $B_{23}$ and
$B_{01}$ can be nonzero. 
General fluctuations of the near-horizon
background \solnear\ or \snn\ will be coupled in a complicated system
of differential equations.
A simple equation is obeyed by the graviton fluctuation $h_{01}$.
This is associated to the
energy-momentum tensor component $T_{01}$ of the Yang-Mills theory.
Let us set all other fluctuations to zero and define 
$\varphi = g^{00} h_{01}$. 
One can check that all equations of motion are satisfied provided
\eqn\jkl{
{e^{2\phi}\over \sqrt{g}}\p_\mu \sqrt{g} e^{-2\phi }g^{\mu\nu}\p_\nu
\varphi =0\ ,
\ \ \ \ \ \ \
\p_0 \varphi =\p_1 \varphi=0\ ,
}
where $g_{\mu\nu}$ is the string-frame metric \solnear\ or \snn .
That is, the equations of motion are satisfied for fluctuations
which are independent of $x_0,x_1$ and obey the {\it scalar} Laplace
equation. In principle one could consider a situation where the fields
depend also on $x_0,x_1$ but the equations might be more complicated.
We will thus consider the gauge theory correlator
$\langle {\cal O} (k){\cal O}(-k)\rangle $, where ${\cal O }(k)$ is the
Fourier transform
of $T_{01}(x)$.


Correlation functions in the gauge theory are obtained by
the usual prescription
\refs{\GKP,\witten}
\eqn\hhdd{
\langle \ \exp \bigg[ \int d^4 k \ \varphi_0(k) {\cal O}(k) \bigg] \ \rangle=
\exp \left[-S_{\rm sugra}\big[ \varphi(k,u)\big] \right]\ ,\ \ \ \ \
i=0,1,2,3,
}
where $\varphi(k,u)$ is the unique solution with $\varphi(k,u)\to \varphi_0(k)$
at some boundary at $u=\Lambda $. We work in momentum space because 
operators in the field theory are defined most naturally in momentum 
space.
%
%
The equation of motion for $\varphi $ in the background \solnear\ then
becomes
\eqn\wwq{
{1\over u^5}\p_u u^5\p_u \varphi - k^2 \bigg({1\over u^4} + a^4
\bigg)\varphi=0\ ,
\ \ \ \ \ \ k\equiv |k|=\sqrt{k_2^2+k_3^2} \ .
}
For $u\cong 0$, this reduces to the Laplace equation in
the $AdS_5\times S^5$ background.
If $k^2 \neq 0$, the
behavior of the solutions at large $u$ is different:
in the $AdS_5$ case, one has $\varphi(u)\sim A u^{-4}+B $,
whereas in the case \wwq\ the solution behaves asymptotically
as $\varphi (u)\sim u^{-5/2} e^{\pm k a^2 u}$.

Interestingly, equation \wwq\ is the same
as the equation for a dilaton fluctuation in the extremal D3 brane
metric, with the replacement $-k^2 \to \omega^2 R^4 $, and $a=1/R$.
This differential equation was solved in \GH\ in terms of Mathieu functions.
Indeed, by the following change of variable,
\eqn\chhh{
u^2={1\over a^2} e^{-2z}\ ,\ \ \ \
\varphi (z)= e^{2z}\psi(z)\ ,\
}
the equation \wwq\ becomes
\eqn\rry{
\bigg({\p ^2\over \p z^2} -2 k^2a^2 \cosh (2z) -4\bigg)\psi(z)=0\ ,
}
which is the Mathieu differential equation with $z\to iz$.

Near $z=\infty $ (or $u\sim 0$), the solution which is well-behaved is of
the form
\eqn\iii{
\varphi (z)\cong e^{2z} K_2(ke^z)\to 0 \ \ {\rm for}\ \ z\to\infty\ .
}
For $z\to -\infty $ (or $u\to \infty $), the solution behaves as
\eqn\jjh{
\varphi(z)\cong D(k)\ a^{5/2}  e^{5z/2}\bigg( \ \exp\big[ka e^{-z}\big]
+B(k)\ \exp\big[-{ka e^{-z}}\big] \bigg)\ .
}
where $D(k)$ is a constant that must be determined by boundary conditions.
The case analyzed in \GH\ can be obtained by taking $z \to z + i \pi/2$;
this implies that the coefficient $B$ in our case is the same as the
reflection coefficient in their situation. 
The correlation functions must be renormalized.
The most reasonable way to do this is to impose the boundary 
condition
\eqn\boundcond{
\varphi(k,\Lambda) = {1 \over ( k a^2  \Lambda)^{5/2} } e^{ ka^2 \Lambda} 
\varphi_0(k) ~,~~~~~ \Lambda \to \infty\ ,
}
where $u = \Lambda$ is the cutoff surface 
where we impose the boundary conditions.
The dependence on $\Lambda$ of this boundary condition can be understood
as follows. We would like to renormalize the theory in such a way that
the solution at finite $u$ remains fixed as $\Lambda \to \infty$. The
$\Lambda$ dependence of \boundcond\ ensures that. The factor of $k$
is ultimately introduced to reproduce the correct IR behaviour. 
One can qualitatively understand it as follows. The boundary condition
on $\varphi$ in the $AdS$ case is that $\varphi$ goes to a momentum
independent constant on
the boundary of $AdS$.
In the large $u$ region of the equation \wwq\ the solution has the form
$\varphi \sim u^{-2}\big[a(k) I_2(ka^2u)+b(k)K_2(ka^2u)\big]$;  
the exponentially increasing term of \jjh\ is in the $I_2$ Bessel
function. We
can
see that  the boundary condition 
\boundcond\  implies that this function goes to a constant at the
origin (plus a term going like $1/u^4$ ).

{}From eq.~\boundcond\ we obtain
\eqn\dequ{
D \sim { 1 \over (k a^2)^{5/2} ( 1 + B e^{- 2 k a^2 \Lambda} ) }\ .
}

We evaluate the action integrating by parts and using \iii , \jjh .
We find a boundary term
\eqn\evalac{\eqalign{
S =& {1 \over 2} \int 
d^2 k \varphi(k,u) u^5 \partial_u \varphi(u,k)|_{u = \Lambda } 
\cr
=& {1\over 2}\int d^2 k  {1\over k^5} 
\left[ div(\Lambda, k, a) -2ka^2 B(k)  + \cdots
\right]
\varphi_0(k) \varphi_0(-k) 
}}
where $div(\Lambda, k, a)$ indicates terms that are divergent when 
$\Lambda \to \infty$ and which are substracted away. 
%
The dots indicate 
terms that vanish when $\Lambda \to \infty $. So we find that the 
correlation function is proportional to the coefficient $B$ in \jjh ,
which in turn is the same as the reflection amplitude in 
\GH :
\eqn\wwk{
\langle {\cal O}(k) {\cal O}(k') \rangle \sim { \delta^2 S \over 
\delta \varphi_0(k) \delta \varphi_0(k') } \sim \delta(k+k') {  B(k)
\over (k a^2)^4 } \ . 
}
This result has the correct IR behavior 
$ \langle {\cal O}(k) {\cal O}(k') \rangle \sim k^4 \log k$, but 
we basically put this in when we included the $k$ dependent factor
in \dequ .\foot{There were some errors in the original version
of this paper.  
As a result, our expressions did not have the correct IR behaviour
(contrary to what was stated in the paper).
We thank L. Rastelli and S. Das for pointing
out this. }
Since the renormalization depends on the momentum it is not possible to
go back to coordinate space in an unambiguous fashion. This is 
a reflection of the non-local nature of the
theory in the ultraviolet. The situation is essentially the same as in 
\shol .

To study the UV regime, we need to determine the value of $B(ka)$
at large $k$.
As explained above, by $z \to z + i\pi/2$ in \rry\ this problem 
is converted into the problem of calculating
the 
reflection coefficient for a particle with an energy
greater than the potential barrier.
This can be done by using the WKB approximation \landau .
For $k^2a^2\gg 1$, we find the following result:
\eqn\wwkk{
\langle T_{01}(k)T_{01}(-k)\rangle 
\cong 
\exp\big[ -c |k|a\big]\ ,\ \
} 
where
\eqn\kkbb{
c= \sqrt{2} \int_0^{\pi/2} dx\ \sqrt{\cos(x)}={2\over\sqrt{\pi} }
\ \Gamma(3/4)^2\cong 1.69\ .
}
Thus the correlator vanishes in the UV limit $k^2\to \infty $. 
This is a drastic change with respect to the commutative case.
We should remember, however, that we have performed a momentum dependent
renormalization, so 
there is some ambiguity in this result. 
We only expect ratios of three (or higher $n$) point
functions to two-point functions to be unambiguous.

Let us now discuss what we expect from the point of
view of field theory. The arguments of \filk\ would naively indicate that all calculations
in the large $N$ limit should reduce trivially to the commutative 
counterparts. We should remember that they had obtained some
phases for the external momenta. Defining operators involves 
taking a particular superposition of bare operators.
It is not easy to find the gauge invariant operator
in the non-commutative Yang-Mills theory representing $T_{01}(k)$
(the Fourier transform of ${\rm Tr\ } F * F$ is not gauge invariant).
In the presence of $b$ we would take a different supperposition of
operators and therefore the results could change.

\newsec{Wilson loops}

Let us first consider the case when only $B_{23}\neq 0$.
The near-horizon geometry is given by \solnear .
A single string stretching from the horizon to the boundary has
the same energy as in the $AdS$ case, i.e. the energy grows with
the cutoff in the same manner. 
Now suppose we bring two of these strings close to each other, at different
points in the $x_2,x_3$ plane.
Naively one would imagine that they connect and give rise to some potential.
This is not what happens. 
The Nambu-Goto 
action for a single string with $x_0=\tau$, $u = \sigma$, $x^i(u)$, $i=2,3$,
is given by 
\eqn\actu{
S = { R^2 \over 2 \pi} 
\int dx_0 du \sqrt{ 1 + (\partial_u x^i)^2 u^4 \hat h}\ .
}
Expanding \actu\ for small fluctuations $x^i$ at large $u$ we get 
$S \sim \int dx_0du(\partial_u x^i)^2 $. This implies that we cannot fix the 
position of the string at infinity since any small perturbation 
implies that $x^i$ grows linearly with $u$. 
We can however specify the slope $k^i$ which the string has as it approaches 
infinity, $x^i \sim k^i u$.
We will see below that, in a loose sense, we can interpret $k$ to be related
to some kind of 
Fourier transform of the Wilson loop. The Wilson loop operator, which 
starts extended along the time direction and seems localized in $x^2,x^3$,
is now replaced by a new operator which is characterized by 
the ``momentum'' $k$ in the $x_2,x_3$ directions. 
In this way we can calculate the potential energy between a quark with 
``momentum'' $k$ with an antiquark with ``momentum'' $k'$. This will vanish
unless $k^i = -{k'}^i$.
This can be seen by taking the action \actu\ and noticing that, 
interpreting $u$ as ``time'' then the canonical momenta conjugate to 
$x^i$ are conserved. So we can choose $k$ to point in the direction $3$.
Denoting by $u_0$ the coordinate of closest approach to the horizon we 
find that 
\eqn\fux{
x^3(u) = \int_{u_0}^u du' { u_0^2 (1
+ a^4 {u'}^4) \over {u'}^2 \sqrt{ {u'}^4 -u_0^4} }\ .
}
{}From the large $u$ behavior we find that $ k = u_0^2 a^4 $. 
We can calculate the energy, which is divergent. The divergent piece
depends on $k$, as we found for the correlation functions. We subtract it
and we get a finite answer which is 
\eqn\enerp{ E = - {R^2 \sqrt{2 \pi} \over \Gamma(1/4)^2}
\sqrt{ 1 + {k^2\over a^4} }{ \sqrt{k} \over a^2} \ .
}
In order to make contact with the standard $AdS$ expression notice that 
for small $k$ then $u_0$ will be very small ($a u_0 \ll 1$). 
Looking at \fux\ we see that we get a large region in the radial coordinate
where we can ignore the term proportional to $a^4{u'}^4$.
In that case we recover
the $AdS$ expression for the coordinate $x$, and in particular we
find
that the separation is $L \sim 1/u_0 \sim a^2 /\sqrt{k}$. 
Substituting in \enerp\ we recover the $AdS$ expression, $E \sim 1/L$.
We did not get this relationship between distance and ``momentum'' 
by performing a Fourier transformation, and it is not standard, 
that is the reason for the 
quotation marks in ``momentum''.

Now consider a string
configuration in the geometry \snn\ (in the particular case $a=a'$)
of the form
\eqn\xxc{
x_1=\tau\ ,\ \ \ \ x_2=\sigma \ ,\ \ \ \ u=u(\sigma )\ ,\ \ \ x_0=x_3=0\ ,
}
which is placed at some given angle in the 5-sphere.
Using eqs.~\solnear ,~\xxc , the action takes the form
\eqn\qqp{
S={ R^2 \over 2\pi }\int d\tau d\sigma 
\sqrt{ \hat h }\sqrt{ (\p_\sigma u)^2+ u^4 \hat h }\ ~,~~~~~~~~
\hat h={1\over 1+a^4u^4}\ .
}
The solution that minimizes the action is given by
\eqn\eeu{
{ u^4\hat h^{3/2}\over \sqrt{ (\p_\sigma u)^2+ u^4 \hat h} }
={ e \over a^2}={\rm const. }\ \ ,
}
or 
\eqn\ddw{
e \p_\sigma u= u^2\sqrt{\hat h} \sqrt{ {a^4 u^4{\hat h}^2 -e^2}} \ ,
}
where $e$ is an integration constant.
The equation \ddw\ is symmetric under $u\to 1/(a^2 u)$ since the 
string metric has this symmetry. We have solutions for $0<e\leq 1/2$.
If $e=1/2$ the solution is a straight worldsheet sitting at $u=a$, 
which is the maximum of the potential that the string sees.
For $0<e<1/2$ the solution oscillates
between $u_{\rm min}$ and $u_{\rm max} = 1/(a^2u_{\rm min})$ 
as we move in $x_2= \sigma$. $u_{\rm max}, u_{\rm min}$ are 
solutions of $\partial_\sigma u =0$. The strings
never get 
to the boundary. If we compactify the $x_1,x_2$ directions
these solutions describe finite action worldsheet instantons. In this case
$e$ will be quantized so that we have an integral number of oscillations
in the $x_2$ circle. 
When $e\ll 1/2$ we get a string configuration which in the region
$u\ll 1/a$ looks like the Wilson loops of $AdS_5$.


\newsec{Generalizations}

\subsec{ D1-D5 system with a $B$ field}

A D1-D5 system with a $B$ field
can be obtained likewise by U-duality.
We start with the usual D1-D5 system in Euclidean space,
with world-volume coordinates $x_0,x_1$, and
make a T-duality transformation in the direction
$x_1$, obtaining a D0-D4 bound state with an extra translational isometry in
$ x_1$.
Then we redefine coordinates as follows,
\eqn\aae{
x_0=x_0'\cos\theta + x_1'\sin\theta \ ,\ \ \ \
\tilde x_1= -x_0'\sin\theta+x_1'\cos\theta\ .
}
By applying a T-duality transformation in the $x_1'$ direction,
we find (restoring the labels $x_0,x_1$ for the 2-plane coordinates and
performing a gauge transformations that changes the asymptotic value of $B$)
\eqn\aaf{
\eqalign{
ds^2_{str} &= (f_1f_5)^{-1/2} h(dx_0^2+dx_1^2)+
f_1^{1/2}f_5^{1/2}(dr^2+r^2d\Omega_3^2) 
\cr
&+ f_1^{1/2}f_5^{- 1/2} dy_ndy_n \ ,\ \ \ \ \ n=1,2,3,4\ ,
\cr
h^{-1}&= (f_1f_5)^{-1} \sin^2\theta+\cos^2\theta \ ,
\ \ \ \ \ \ f_{1,5}=1+{\ap R^2_{1,5}\over r^2}\ ,
\cr
e^{2\phi } &= g^2 {f_1\over f_5} h\ , \cr 
\chi=& i {1\over g} \sin\theta \ f_1^{-1}\ ,\ \ \ \ 
A_{y_1y_2y_3y_4}=i {1\over g} \sin\theta f_5^{-1} \ , \cr
B_{01} &= {\sin\theta\over \cos\theta}\ (f_1f_5)^{-1} h\ ,
\cr
dA_2 &={1\over g}
\cos\theta \ \big(i \ d(h f_1^{-1}) \wedge dx_0\wedge dx_1
+ 2 R^2_5 \epsilon_3 \big)\ ,
}}
where $\epsilon_3$ is the volume element of a unit radius 3-sphere.
By a similar rescaling of parameters as in \limits\ one obtains the
near-horizon geometry
\eqn\aah{
ds^2_{str}=\alpha 'R^2\bigg[u^2 \hat h 
(dx_0^2+dx_1^2)+{du^2\over u^2}+
d\Omega_3^2\bigg]\ + {R_1\over R_5} dy_ndy_n\ ,
\
}
\eqn\aai{ \eqalign{
e^{2\phi } &= {\hat g}^2 \hat h\ ,\
\ \ \ \ 
\hat h={1\over 1+a^4u^4}\ ,\ \ \ \ 
\cr
B_{01} &= B_\infty {a^4u^4 \over 1+a^4u^4}\ ,\ \ \ \ B_\infty =\alpha
'{R^2\over a^2}\ ,
\cr
\chi &={i\over \hat g} a^2 u^2\ ,\ \ \ \ \ 
A_{y_1y_2y_3y_4}={i\over \hat g}{R_1^2\over R_5^2} a^2 u^2\ ,
\cr
dA_2 &={1\over \hat g} {R_1 R_5} \bigg( i \alpha ' 
d(u^2 \hat h)\wedge dx_0\wedge dx_1+2 \epsilon_3 
\bigg)\ ,
\cr
R^4 & =R_1^2R_5^2\ ,\ \ \ \ {a^2\over R^2}=\alpha ' \tan\theta\ ,
\ \ \ \ \ \hat g={g\over\alpha'} {a^2\over R_5^2}\ 
.
}}
Near $u=0$ and near $u=\infty $, the metric approaches $AdS_3\times
S^3\times T^4$.
Although the string frame metric is symmetric under $u\to 1/(a^2u)$,
the dilaton and gauge fields behave differently in the UV region $u=\infty
$. All these solutions, \aaf \aah , preserve 8 supersymmetries.

A D1-D5 configuration with $B$ components in two five-brane directions
$y_1, y_2$ which are transverse to the D1-brane can be obtained
as follows. We apply a T-duality in $y_2$, do a rotation in the $y_1, ~y_2$
plane, a T-duality back on $y_2$ and a gauge transformation to change
the asymptotic value of $B$. 
The solution is given by
\eqn\eew{
\eqalign{
ds^2_{str}=& f_1^{-1/2}f_5^{-1/2}\big(-dx_0^2+dx_1^2\big) 
+f_1^{1/2}f_5^{1/2}\big( dr^2+r^2d\Omega_3^2 \big)+ 
\cr
&+f_1^{1/2}f_5^{-1/2}\big[ h(dy_1^2+dy_2^2) + dy_3^2+dy_4^2\big]
\cr
e^{2\phi}&= g^2{f_1 \over f_5} h\ ,\ \ \ \ \ 
B_{y_1y_2}= {\sin\theta\over \cos\theta } {f_1\over f_5} h\ ,
\cr
h^{-1} &=\cos^2\theta +\sin^2\theta {f_1\over f_5}\ ,
\cr
dA_2 &={1\over g}\cos\theta\ \big( df_1^{
-1}\wedge dx_0\wedge dx_1 + 2 R_5^2\epsilon_3 \big)\ ,
\cr
F= & G + *_{10}\ G \ , \ \ \ \ 
G = \sin \theta \ h { f_1 \over f_5} \ dy_1\wedge dy_2\wedge \big(
df_1^{-1} \wedge dx_0 \wedge dx_1 + 2 R_5^2 \epsilon_3 \big) \ .
}}

General D$p$ brane solutions with a B-field can be constructed
in the same way that \solmet\ was obtained.
The resulting background is given by
\eqn\solmet{\eqalign{
ds^2_{str} &= f^{-1/2} [ - dx_0^2 + dx_1^2 +{h}(dx_2^2 + dx_3^2 )
+...+dx_p^2]
+ f^{1/2} ( dr^2 + r^2 d \Omega^2_{8-p} )\ ,
\cr
f = & 1 + {R^{7-p} \over r^{7-p} } ~, ~~~~~~~~~
h^{-1} =\sin^2\theta f^{-1}+\cos^2\theta \ ,
\cr
B_{23} =& {\sin\theta\over \cos\theta }\ \ f^{-1} h \ ,
\cr
e^{2\phi} &= g^2 f^{3-p\over 2} h \ .
}}
And we also have RR fields excited corresponding to D$p$ brane charges and
D$(p-2)$ brane charge densities. 
It is interesting to note that the string coupling decreases faster
at infinity than in the commutative case. This suggests that maybe
it is possible to define the non-commutative version of the D6 brane
when we have two $B$ fields\foot{This issue was raised by S. Kachru.}.
By taking the large $N$ limit of this field
theory we would describe M-theory on $T^6$ according to \BFSS .
{}From our count of
degrees of freedom above 
it looks like we will have problems similar to the ones 
appearing in the commutative
case. In particular, we can calculate the entropy which gives the
same result as in the commutative case. The entropy goes
as $S \sim E^{3/2} $, which implies that the system has a negative specific
heat. This does not obviously imply non-decoupling from the bulk 
and a more detailed analysis is necessary.

\subsec{ M5 brane with $C$-field}

One can similarly find an M5 brane solution with a $C$-field.
In this case we get
\eqn\mfive{\eqalign{
ds^2_{11} =& k^{1/3} f^{1/3} \big[ {1\over f}
\big( -dx_0^2 + dx_1^2 + dx_2^2 \big)
+ {1\over k}\big(dx_3^2 + dx_4^2 + dx_5^2 \big) + dr^2 + r^2 
d\Omega_4^2 \big]\ ,
\cr
f = & 1 + { R^3 \over r^3 } ~,~~~~~~~
k=\sin^2\theta+\cos^2\theta \ f\ ,
\cr
dC_3 =& \sin\theta\ df^{-1}\wedge dx_0\wedge dx_1\wedge dx_2+
\cos\theta \ 3 R^3 \epsilon_4 
-6\tan\theta\ d(k^{-1})\wedge dx_3\wedge dx_4\wedge dx_5\ .
}}
This solution appeared in \RT\ (eq. (2.26)~) and
it was interpreted as a 2-brane lying within a
5-brane. 
One can similarly introduce a decoupling limit in such a way that 
we obtain the solution \mfive\ but with $f \to R^3/r^3$. 
Notice that the fivebrane charge is given by $ \pi N = R^3/\cos\theta $.
The DLCQ definition of this theory was considered in \ABS .

\medskip

\bigskip\bigskip


\centerline{\bf Acknowledgments}

We would like to thank N. Nekrasov, N. Seiberg and E. Witten for 
interesting discussions.
We thank L. Rastelli and S. Das for pointing out errors in section 4
in the previous version of this paper. 
JM is supported by DOE grant
DE-FG02-92ER40559 and the Sloan and Packard
foundations. JM also wants to thank the Raymond and Beverly Sackler
fellowships and the 
Institute for Advanced Study where this research was started. 
JR would like to thank the support of Conicet and 
Fundaci\' on Antorchas, and CERN and ICTP for hospitality.

\listrefs

\bye